\documentclass[onecol]{epl2}

\title{Photonic analogue of Josephson effect in a dual-species optical-lattice cavity}

\author{Soi-Chan Lei\inst{1}, Tai-Kai Ng\inst{2}, \and Ray-Kuang Lee\inst{1,3}}
\institute{
\inst{1} Department of Physics, National Tsing-Hua University, Hsinchu 300, Taiwan\\
\inst{2} Department of Physics, Hong Kong University of Science and Technology, Hong Kong\\
\inst{3} Institute of Photonics Technologies, National Tsing-Hua University, Hsinchu, Taiwan}


\abstract{We extend the idea of quantum phase transitions of light in the photonic Bose-Hubbard model with interactions to two atomic species by a self-consistent mean field theory.
The excitation of two-level atoms interacting with coherent photon fields is analyzed with a finite temperature dependence of the order parameters.
Four ground states of the system are found, including an isolated Mott-insulator phase and three different superfluid phases.
Like two weakly coupled superconductors, our proposed dual-species lattice system shows a photonic analogue of Josephson effect.
The dynamics of the proposed two species model provides a promising quantum simulator for possible
 quantum information processes.}

\begin{document}

\maketitle
Driven by quantum fluctuations at absolute zero temperature, quantum phase transitions (QPTs) have been intensively studied in interacting many-body problems \cite{book}.
Typically, it is difficult to control and probe such exotic quantum phenomena for strongly correlated electronic systems in condensed matter physics.
Optical lattices - artificial crystals made by interfering laser beams - offer a versatile platform for studying the QPTs of trapped Bose gases \cite{Greiner02}.
In this situation, one can describe the many-body dynamics from a Mott-insulator (MI) phase to a superfluid (SF) phase in a gas of ultracold atoms with periodic potentials by using the Bose-Hubbard model that includes an on-site two-atom interaction and hopping between adjacent sites \cite{Jaksch98}.

Instead of weakly interacted ultracold atoms, photons are non-interacting bosons, and in purely photonic systems there is no possibility to have any QPTs.
For a pure Bose system, the conducting phase at zero temperature is presumably always superfluid \cite{Fisher89}.
However, engineered composites of optical cavities, few-level atoms, and laser lights can form a strongly interacting many-body system to study the concepts and methods in condensed matter physics from the viewpoint of quantum optics.
In this case, a photonic condensed matter analogue could be realized with state-of-the-art photonic crystals embedded with high-Q defect cavities.
Like optical lattices for matter waves, the quantum phase transitions of photonic insulator (excitations localization) to superfluid (excitations delocalization) are predicted by the Bose-Hubbard model with additional photon-atom interactions \cite{Greentree06, Hartmann06,  Angelakis06}.
Related quantum transitions also have been predicted in a Heisenberg spin $1/2$ Hamiltonian \cite{spin07}, two species Bose-Hubbard model \cite{polariton07}, and solved exactly in the one dimensional case \cite{glassy07}.
This opened the possibilities to study critical quantum phenomena in conventional condensed matter systems by manipulating the interaction between photons and atoms.

\begin{figure}
\includegraphics[width=8.0cm]{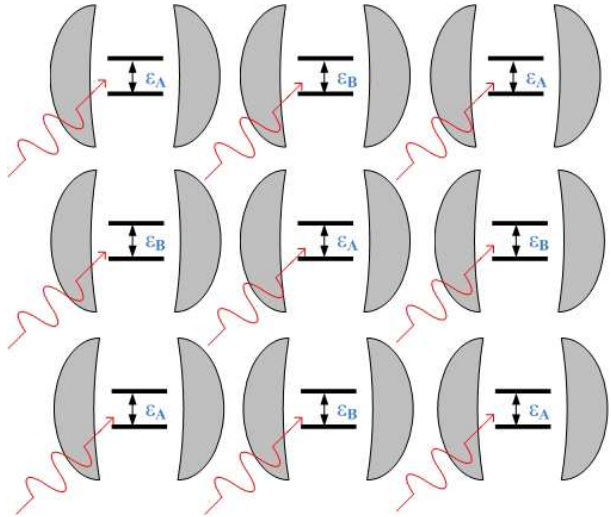}
\caption{Illustration schematic for the proposed system.
An array of high-Q electromagnetic cavities is formed in the configuration of the square lattice, and each cavity contains a single two-level atom of the type  $A$ or $B$, which is spaced at intervals. In this configuration, the center cavity has four nearest inter-species neighbors (different atomic type) and four next-nearest intra-species neighbors (the same atomic type). $\epsilon_A$ and $\epsilon_B$ are the transition energies for the atomic species $A$ and $B$, respectively. The incident optical field interacting with the bipartite lattice is also shown in the red color.}
\label{Fig:1}
\end{figure}

Recently, we illustrate the generality of the method by constructing the dressed-state basis for an arbitrary number of two-level atoms (TLAs) \cite{SoiChan08}.
As the number of TLAs increases, collective effects due to the interactions of atoms among themselves give rise to intriguing many-body phenomena.
With the Dicke-Bose-Hubbard Hamiltonian, we show that the Mott insulator to superfluid QPTs with photons can be realized in an extended Dicke model for an arbitrary number of two-level atoms.
As quantum many-body effects associated with Bose-Einstein condensations in optical lattices have been reported for a long-time \cite{Giamarchi08}, it is believed that the realization of strongly correlated many-body photonic cavity quantum electrodynamic(QED) systems in experiments is also likely to happen soon.

From the view point of a practical experimental setup, instead of modifying the Q-value of individual optical cavities, it is more easier to allocate different atomic species in different photonic cavity sites.
More recently, based on scanning electron microscopy technologies, the addressability in a two-dimensional optical lattice for ultracold atoms has been demonstrated experimentally down to the single-atom and single-site level \cite{single-site}.
A natural consequence of the studies on the QPT of light is to consider the atom-light interacting system with different atomic species.
In this work, we investigate theoretically QPTs of light from the MI to SF quantum phase transitions in high Q-value optical cavities with two species of atoms based on the  photonic Bose-Hubbard model embedded with the Jaynes-Cummings interaction Hamiltonian.
A rather rich low temperature phase diagram emerges even for a relatively simple structure of the cavity lattice.
As the phase diagram of two-component bosons on an optical lattice \cite{Altman08,Mishra08}, we show that separated phases corresponding to MI and composite SF exist.
Moreover, these composite SF phases have an analogue to the well known Josephson effect in a two weakly coupled superconductor.
We expect that more controllable light-wave technologies should lead to an enhanced understanding of the QPTs of light with distinctive properties, improvements in the organization of the ground-state wave function, and the introduction of new applications.

For two atomic species interacting with photons, we consider an array of two-dimensional square photonic bandgap micocavities with two types of two-level atoms (TLAs) labelled as $A$ and $B$, as illustrated in Fig.\ref{Fig:1}.
The array is made by high-Q electromagnetic cavities.
Each cavity contains a single TLA of the type $A$ or $B$, which is spaced at intervals.
The atoms are assumed to interact strongly with photons in the photon-blockade regime so that a coherent composite polaritonic (photon and atom) system is formed in the presence of radiation fields \cite{Greentree06, SoiChan08}.
Photons can travel from one cavity to another.
We assume that travelling process for photons is dominated by the hopping between adjacent sites.
The value of the hopping matrix elements is a function of the distance between the cavities, which is a tunable parameter in our system.

The interaction between a two-level atom and the quantized photon mode of an optical cavity  is described by the Jaynes-Cummings model \cite{Jaynes63}, with the two-site Hamiltonian for the $i$-th unit cell (two cavities numbered as $2i$ and $2i+1$ for $A$- and $B$-type atoms, respectively),
\begin{eqnarray}
\label{eqJCM}
H^{two-site}_i &=& \omega\, a^\dagger_{2i}a_{2i}+\epsilon_{A,2i}\sigma^z_{A,2i} + g_{A,2i}(a_{2i}\sigma^\dagger_{A,2i}+a^\dagger_{2i}\sigma^-_{A,2i}),\nonumber\\
&+& \omega\, a^\dagger_{2i+1}a_{2i+1} + \epsilon_{B,2i+1}\sigma^z_{B,2i+1} + g_{B,2i+1}(a_{2i+1}\sigma^\dagger_{B,2i+1}+a^\dagger_{2i+1}\sigma^-_{B,2i+1}),
\end{eqnarray}
where $\epsilon_{A(B), i}$ is the transition energy for the TLA in  $i$-th cavity and $\omega$ is the radiation field frequency which we have assumed to be the same at all cavities.
The atom-photon coupling, $g_{A(B),i}$, is assumed to be real here.
$a^\dagger_i$ and $a_i$ are the raising and lowering operators for photons, respectively.
$\sigma^\dagger_{A(B), i}$ excites the $A$(or $B$)-type two-level atom from the low energy state to high energy state and $\sigma^-_{A(B),i}$ describes a reverse process.
By including the hopping process for photons, the total Hamiltonian describing our proposed system for a configuration of $2 N$ cavities is
\begin{eqnarray}
\label{eqFJCH}
H=-\sum_{\langle i,j\rangle}^{2N}\kappa_{i,j}{a^\dagger_i}{a_j}+\sum_{i=1}^N \left[-\mu (n_{2i}+n_{2i+1})+H^{two-site}_i\right],
\end{eqnarray}
where $\kappa_{i,j}$ is the hopping matrix element and $i,j=1,\dots,2N$ are the cavity indices.
We have also introduced a chemical potential term $\mu$ for photons which control the overall strength of the photon field, and $n_i=a^\dagger_ia_i$ is the photon number operator.
For simplicity, in this configuration, each cavity has four nearest inter-species neighbors (different atomic type) with the hopping coefficient $\kappa_{A\leftrightarrow B} = \kappa_{B\leftrightarrow A} \equiv \kappa$ ,
 and four next-nearest intra-species neighbors (the same atomic type) with the hopping coefficient $\kappa_{A\leftrightarrow A} = \kappa_{B\leftrightarrow B} \equiv \kappa^\prime$.

To solve the model Hamiltonian, we introduce a self-consistent mean field approach by approximating 
\begin{eqnarray}
\label{mfd}
a \sigma^{\pm}_{A(B)} &\approx& \langle a\rangle \sigma^{\pm}_{A(B)} + a\langle \sigma^{\pm}_{A(B)} \rangle - \langle a \rangle \langle \sigma^{\pm}_{A(B)} \rangle.
\end{eqnarray}
A similar decomposition is also applied to the terms associated with $a^\dagger \sigma^{\pm}_{A(B)}$.
It is convenient to introduce two superfluid order parameter $\psi_{A(B)}$ for photons and two TLA order parameter $J_{A(B)}$ for the atomic species $A$ and $B$, respectively, where
\begin{eqnarray}
\label{eqSOP}
&& \psi_{A} \equiv \langle a_{2i} \rangle \equiv \langle a^\dagger_{2i} \rangle,\\
&& \psi_{B} \equiv \langle a_{2i+1} \rangle \equiv \langle a^\dagger_{2i+1} \rangle,\\
&& J_{A(B)} \equiv  \langle \sigma^{\dagger}_{A(B)} \rangle \equiv \langle \sigma^{-}_{A(B)} \rangle.
\end{eqnarray}
The order parameter introduced here is indeterminate at the point of the phase transition in
our photonic Bose-Hubbard Hamiltonian \cite{Greentree06, Hartmann06, Angelakis06}.
The system is in the photonic superfluid phase for a non-zero order parameter $\psi_{A(B)} \neq 0$, while in the insulator phase for a zero order parameter $\psi_A = \psi_B = 0$.
And $J_{A(B)}$ can be viewed as an order parameter for the atomic coherent states.
The physical meaning for $J_{A(B)}$ as a coherent state order parameter for the atoms lies on the coherent superposition of the ground and exacted state of the atoms.
When the system temperature goes down, the coherent superposition of the ground and exacted state of the atoms becomes larger, from Eq. (\ref{hfd2}), and it reaches in its largest value at absolute zero temperature as expected.
With these order parameters, the possibility to have QPT at low temperature would be demonstrated in the right parameter regimes.

In the following we assume that both $\psi_{A(B)}$ and $J_{A(B)}$ are real numbers and spatially independent, i.e. no spontaneous symmetry-breakings in the translation and rotation for the infinite system considered here.
The assumption that $\psi_{A(B)}$ and $J_{A(B)}$ are real numbers will be justified later by the
self-consistent mean-field solution.
In general it is very difficult to diagonalize the model Hamiltonian in Eq. (\ref{eqFJCH}) and find out all the desired QPTs of the system.
However, it may be instructive to diagonalize part of the total Hamiltonian first.
With the mean-field approximation, Eq.(\ref{eqFJCH}) can be decomposed into two Hamiltonians, labelled as $H^{f}$ and $H^{p}$, where the atomic and photonic operators can be considered separately.
For one atom ($A$ or $B$) inside each cavity, the mean-field Hamiltonian for atomic operators is
\begin{eqnarray}
\label{hf}
H^{f}_{A(B)} &=& \epsilon_{A(B)} \sigma^z_{A(B)} + g_{A(B)}\psi_{A(B)}\left[\sigma^{\dagger}_{A(B)}+\sigma^{-}_{A(B)}\right].
\end{eqnarray}
This Hamiltonian can be diagonalized easily in the form
\begin{eqnarray}
\label{hfd}
H^{f}_{A(B)}=E_{A(B)}\left[\frac{1}{e^{\beta E_{A(B)}}+1}-\frac{1}{e^{-\beta E_{A(B)}}+1}\right],
\end{eqnarray}
with the eigen-energy
\begin{eqnarray}
\label{eegy}
E_{A(B)}={\{}\left[g_{A(B)}\psi_{A(B)}\right]^2+\epsilon_{A(B)}^2{\}}^{1/2},
\end{eqnarray}
Then the TLA order parameter $J_{A(B)}$ can be derived as
\begin{eqnarray}
\label{hfd2}
J_{A(B)}  =  - \frac{\psi_{A(B)} g_{A(B)}}{2 E_{A(B)}}\, Tanh[\frac{E_{A(B)}}{2 T}],
\end{eqnarray}
where we introduce the temperature for the system, $T$, by setting $\beta = 1/k_B T$ with the Boltzmann constant $k_B = 1$.
Note that here $J_{A(B)}$ is a real number, and
 consistent with our initial assumption. From Eq. (\ref{hfd2}), it can be seen clearly that only at zero temperature, the  TLA order parameter $J_{A(B)}$ has a maximum value.

For the radiation fields, the related photonic mean-field Hamiltonian in Eq.(\ref{eqFJCH}) can be separately for the two sites $2i$ and $2i+1$ as
\begin{eqnarray}
\label{pmf}
H^p_{2i} &=& -\sum_{\langle i,j\rangle}^{N}\kappa_{i,j}{a^\dagger_i}{a_j} + \sum_{i=1}^N \left[(\omega-\mu )n_{2i}+g_{A, 2i} J_{A}(a_{2i}+a^\dagger_{2i})\right],\\
H^p_{2i+1} &=& -\sum_{\langle i,j\rangle}^{N}\kappa_{i,j}{a^\dagger_i}{a_j} + \sum_{i=1}^N \left[(\omega-\mu )n_{2i+1}+g_{B, 2i+1} J_{B}(a_{2i+1}+a^\dagger_{2i+1})\right].
\end{eqnarray}
These Hamiltonians can be diagonalized by the Fourier transforms for the even- and odd-numbered sites, i.e.
\begin{eqnarray}
a_{A,k} &= & \sum_{i}^N e^{j\vec{k}\cdot\vec{r}_{2i}}a_{A, 2i},\\
a_{B,k} &= & \sum_{i}^N e^{j\vec{k}\cdot\vec{r}_{2i+1}}a_{B, 2i+1}.
\end{eqnarray}
In terms of $a_{A(B), k}$, the combined photonic mean-field Hamiltonian of two sites in a unit cell becomes
\begin{eqnarray}
\label{pmff}
H^{p} & = &  H^{p}_{2i} + H^{p}_{2i+1},\\ \nonumber
& = & [g_A J_A (a^\dagger_{A, k=0} + a_{A, k=0}) + g_B J_B (a^\dagger_{B, k=0} + a_{B, k=0})], \\ \nonumber
&+& \sum_{k}\Omega_0(\vec{k})\left[a^\dagger_{A, k} a_{A,k} + a^\dagger_{B, k} a_{B, k}\right] + \sum_k\Omega_1(\vec{k})\left[a^\dagger_{A, k} a_{B, k}+a^\dagger_{B,k} a_{A,k}\right],
\end{eqnarray}
where
\begin{eqnarray}
\Omega_0(\vec{k}) &=& -4 \kappa^\prime\, Cos(k_x)\, Cos(k_y)+\omega-\mu,\\
\Omega_1(\vec{k}) &=& -2 \kappa\left[Cos(k_x)+\, Cos(k_y)\right],
\end{eqnarray}
with the Fourier vector $\vec{k} = k_x \hat{x} + k_y \hat{y}$.
The assumption that $J_{A(B)}$ is spatially independent implies that the atoms couple only to the $\vec{k}=0$ photon mode.
In other words, we only search for the spatially homogeneous solutions in our mean-field theory.

The photonic mean-field Hamiltonian $H^p$ in Eq. (\ref{pmff}) can be diagonalized by introducing symmetric and antisymmetric photon field operators $a_{sym, k}$ and $a_{ant, k}$, i.e.
\begin{eqnarray}
a_{sym, k} \equiv (a_{A,k}+a_{B,k})/{\sqrt{2}},\\
a_{ant, k} \equiv (a_{A,k}-a_{B,k})/{\sqrt{2}}.
\end{eqnarray}
And in terms of $a_{sym, k}$ and  $a_{ant, k}$ field operators, the photonic mean-field Hamiltonian $H^p$ becomes
\begin{eqnarray}
\label{eq8}
H^{p} &=& \sum_k \left[\Omega_{sym}(\vec{k}) a^\dagger_{sym, k} a_{sym, k} +\Omega_{ant}(\vec{k}) a^\dagger_{ant, k} a_{ant, k} \right]\\\nonumber
&+& g_{sym} J_{sym}(a^\dagger_{sym, k=0}+ a_{sym, k=0}) + g_{ant} J_{ant}(a^\dagger_{ant, k=0}+ a_{ant, k=0}),
\end{eqnarray}
where
\begin{eqnarray}
\Omega_{sym}(\vec{k}) &=& \Omega_0(\vec{k})+\Omega_1(\vec{k}),\\
\Omega_{ant}(\vec{k}) &=& \Omega_0(\vec{k})-\Omega_1(\vec{k}),\\
g_{sym} J_{sym} &=& {1\over\sqrt{2}} (g_A J_A+g_B J_B),\\
g_{ant} J_{ant} &=& {1\over\sqrt{2}} (g_A J_A-g_B J_B).
\end{eqnarray}
Here the stability of photon fields requires a non-zero energy (frequency), i.e. $\Omega_{sym(ant)}(\vec{k})\geq 0$.
Equivalently we have the condition $(\omega-\mu)-4\kappa^\prime > 4 \kappa$ required for photon fields.

From the coupling of photon field to atoms, one may imply that the $\vec{k}=0$ mode of the photon develops a ground state, with the expectation value
\begin{eqnarray}
\label{bc1}
&&\langle a_{sym, k=0}\rangle = -\frac{g_{sym}\,J_{sym}}{\Omega_{sym}(\vec{k}=0)},\\
\label{bc2}
&&\langle a_{ant, k=0}\rangle =
-\frac{g_{ant}\,J_{ant}}{\Omega_{ant}(\vec{k}=0)},\\
\label{bc3}
&&\psi_{A}={1\over\sqrt{2}}\left[\langle a_{sym,k=0}\rangle + \langle a_{ant, k=0}\rangle\right],\\
\label{bc4}
&&\psi_{B}={1\over\sqrt{2}}\left[\langle a_{sym,k=0}\rangle - \langle a_{ant, k=0}\rangle\right].
\end{eqnarray}
Eqs. (\ref{bc1}-\ref{bc4}) together with Eq.\ (\ref{hfd2}) are the main results of the present work, which form a self-consistent set of equations determining the photonic superfluid order parameters $\psi_{A(B)}$ and the atomic coherent state order parameter for TLAs $J_{A(B)}$. Based on these results, the proposed two-site Hamiltonian will be used to study the QPT of light analytically and  numerically in the following.

In general, there are more than one solution to the self-consistent mean-field equations.
But by calculating the corresponding free energy, one can determined the equilibrium state which occupies with the lowest free energy.  The free energy density of our system $F_s$ can be written as $F_s=F_f +F_p-E_m$, where $F_f$, $F_p$, and $E_m$ are the free energy densities associated with the mean-field Hamiltonian in Eq. (\ref{hfd}), photonic mean-field Hamiltonian in Eq. (\ref{eq8}), and the correction term to the double counting from our mean-field decomposition in Eq. (\ref{mfd}), respectively, i.e.
\begin{eqnarray}
F_f &=& -E_A - E_B - \frac{2}{\beta}\left[\ln(1+e^{-\beta E_A})+\ln(1+e^{-\beta E_B})\right],\\
F_p &=& {1\over V} \sum_{k}{1\over\beta}ln\left[(1-e^{-\beta\Omega_A(\vec{k})})(1-e^{-\beta\Omega_B(\vec{k})})\right]
- \frac{(g_A J_A)^2}{\Omega_A} - \frac{(g_B J_B)^2}{\Omega_B},\\
E_m &=& 2 \left[ \frac{(g_A J_A)^2}{\Omega_A} + \frac{(g_B J_B)^2}{\Omega_B}\right],
\end{eqnarray}
where $V$ is the volume of the system.

At zero temperature, the ground state for the mean-field Hamiltonian in Eq. (\ref{hfd2}) has the simple relation,
\begin{eqnarray}
\label{fgd}
J_{A(B)}=-\frac{\psi_{A(B)}g_{A(B)}}{2E_{A(B)}},
\end{eqnarray}
and the corresponding ground state energy density per cavity for the system is
\begin{eqnarray}
\label{eg}
E_g={1\over2}\left[\frac{(g_A J_A)^2}{\Omega_A}+\frac{(g_B J_B)^2}{\Omega_B}-E_A-E_B\right].
\end{eqnarray}

Combine the mean-field equations in (\ref{bc1}-\ref{bc4}) and Eq. (\ref{hfd2}), one can obtain $\psi_{A(B)}$ in the following matrix form,
\begin{eqnarray}
\label{mmf}
\left(
\begin{array}{c}
\psi_A \\ \psi_B
\end{array}
\right)=\left(
\begin{array}{cc}
\Omega_+^{-1} & \Omega_-^{-1} \\
\Omega_-^{-1} & \Omega_+^{-1}
\end{array}
\right)\times\left(
\begin{array}{c}
{g_A^2\over 4E_A}\, {Tanh}({E_A\over 2T})\psi_A  \\
{g_B^2\over 4E_B}\, {Tanh}({E_B\over 2T})\psi_B
\end{array}
\right)
\end{eqnarray}
where $\Omega_{+}^{-1}=\Omega_{sym}^{-1}+\Omega_{ant}^{-1}$ and $\Omega_{-}^{-1}=\Omega_{sym}^{-1}-\Omega_{ant}^{-1}$ are the related photon dispersions.
Eq. (\ref{mmf}) supports an trivial solution to the mean-field equations, i.e. $\psi_A=\psi_B=0$.
But non-trivial solutions can be obtained only numerically in general for the reason that  $E_{A(B)}$ is a function of $\psi_{A(B)}$ in Eq. (\ref{eegy}).

To give a simple picture for the QPTs of light in our system, we first consider the special case when the atoms $A$ and $B$ are identical.
For this case $\psi_A=\psi_B=\psi$,  $E_A=E_B=E$, $\epsilon_A=\epsilon_B=\epsilon$, and $g_A=g_B=g$, we can derive the photonic superfluid order parameter explicitly,
\begin{eqnarray}
\label{sol1}
\psi={g^2\over4E\Omega_{sym}}\,{Tanh}({E\over 2T})\,\psi.
\end{eqnarray}
In particular, non-trivial solution, $\psi\neq0$, exists at $T=0$ only when the following inequality is satisfied,
\begin{eqnarray}
{g^2\over 4\Omega_{sym}}>\epsilon,
\end{eqnarray}
and the corresponding order parameter solution is
\begin{eqnarray}
\psi=\sqrt{({g\over4\Omega_{sym}})^2-(\frac{\epsilon}{g})^2},
\end{eqnarray}
which indicates a zero-temperature phase transition of superfluid to Mott-insulation phases at the condition
\begin{eqnarray}
\label{gqpt}
{g^2\over4\Omega_{sym}}=\epsilon.
\end{eqnarray}
A corresponding finite temperature transition also exists at the critical temperature $T_c$ given by
\begin{eqnarray}
{4\epsilon\Omega_{sym}\over g^2} = {Tanh}({\epsilon\over 2T_c}).
\end{eqnarray}

To illustrate what happens when the atoms $A$ and $B$ are different, we consider the limit case of $\Omega_-^{-1}=0$ (or $\kappa=0$), i.e. without inter-species hopping effects.
In this case, at $T = 0$ the atoms $A$ and $B$ couple separately to the radiation fields.
From Eq. (\ref{gqpt}), each of the two species has a MI to SF phase transition at $g_{A}= \sqrt{4\Omega_+\epsilon_{A}}$ and $g_{B}= \sqrt{4\Omega_+\epsilon_{B}}$, respectively.
In such a way, there should be four possible phases in the parameter plane ($g_A$, $g_B$).
These four phases correspond to (1) both of the two-species atoms are in the MI state, $g^2_{A(B)}  <4\Omega_+\epsilon_{A(B)}$; (2) only $A$-type atoms are in the SF state, named as the SF-A state for $g^2_A < 4\Omega_+\epsilon_A$ and $g^2_B > 4\Omega_+\epsilon_B$; (3) the SF-B state with the case for only $B$-type atoms are in the SF state; and (4) all the atoms are in the SF state, i.e. the co-existence SF-AB state for $g^2_{A(B)} > 4\Omega_+\epsilon_{A(B)}$.

To give a qualitative analysis on the possible four phase states, we assume  $\Omega_-^{-1}$ to be small and perform a perturbative expansion on the limit case $\Omega_-^{-1}=0$ for the solutions of the mean-field Hamiltonian in Eq.  (\ref{mmf}) at zero temperature $T=0$.
By expanding the superfluid order parameter  $\psi_{A(B)}=\psi^{(0)}_{A(B)}+\delta\psi_{A(B)}$, to the zero-th order one can again obtain,
\begin{eqnarray}
\label{psi0}
\psi_{A(B)}^{(0)} = \left\{ \begin{array}{l@{\quad ; \quad}l}
0 & \mbox{for} \quad {g^2_{A(B)}\over4\Omega_+}<\epsilon_{A(B)}  \\
\sqrt{\left[{g_{A(B)}\over4\Omega_+}\right]^2-\left[{\epsilon_{A(B)}\over g_{A(B)}}\right]^2} &
\mbox{for} \quad {g^2_{A(B)}\over4\Omega_+}>\epsilon_{A(B)} \end{array} \right. ,
\end{eqnarray}
and  to the first order expansion in $\Omega_-^{-1}$,
\begin{eqnarray}
\label{psi1}
\delta\psi_{A(B)} =   \left\{ \begin{array}{l@{\quad ; \quad}l}
{g_{{B(A)}}^2\over4\left[1-{g_{{A(B)}}^2\over4\epsilon_{{A(B)}}\Omega_+}\right]\Omega_-E_{B(A)}^{(0)}}
   \psi_{B(A)}^{(0)} &
\mbox{for} \quad \psi_{A(B)}^{(0)}=0\\
{\Omega_+\over\Omega_-}{g_{B(A)}^2E_{A(B)}^{(0)2}\over g_{A(B))}^4E_{B(A)}^{(0)2}} {\psi_{B(A)}^{(0)}\over\psi_{A(B)}^{(0)3}}; &
\mbox{for}\quad \psi_{A(B)}^{(0)}\neq0 \end{array} \right. ,
\end{eqnarray}
with $E_{A(B)}^{(0)}=\sqrt{g_{A(B)}^2\psi_{A(B)}^{(0)2}+\epsilon_{A(B)}^2}$.

With above perturbative results, one can have a clear physical interpretation for the QPT in our proposed system.
When all the atoms are in the MI phase, $\psi_{A}^{(0)}=\psi_B^{(0)}=0$, both of the perturbed superfluid order parameters $\delta\psi_{A}$ and $\delta\psi_{B}$ are zero as expected.
The MI phase of the system is not modified by the perturbation in $\Omega_-^{-1}$.
On the contrary, the properties of SF states are however strongly modified.
From the second line in Eq. (\ref{psi1}), one can easily find that $\delta\psi_{A(B)}$ is non-zero as  long as  $\psi_{B(A)}^{(0)} \neq0$, which is independent of whether $\psi_{A(B)}^{(0)}$ is zero or not. 
In such a scenario, one of atomic species in the superfluid phase can induce a non-zero superfluid order parameter on the other species of atoms, which is originally in the MI state.
Unlike the case for only one species of atoms, in our system both species of atoms develop nonzero order parameters with the superfluidity driven by another type of atomic species, similar to the Josephson effect in the two weakly coupled superconductors \cite{Josephson}.

The coupling between two atomic species through the radiation fields smears out the difference between the three superfluid phases mentioned above, i.e  SF-A, SF-B, and the coexistence SF-AB phases, and turns the phase transitions between these different phases into crossovers. Nevertheless the qualitative properties of these three SF states are quite different.
For example, when turning on an additional laser with the frequency $\sim\epsilon_B$ on our system, the input light resonates strongly with the $B$-type atoms, and the superfluidity of $A$-type atoms would be sequentially modified depending on their original state of the system.
In an original SF-B phase, superfluidity of $A$-type atoms would be destroyed because of decoupling from $B$-type atoms, while an original co-existence SF-AB phase would be driven into the SF-A phase.

\begin{figure}
\includegraphics[width=8.0cm]{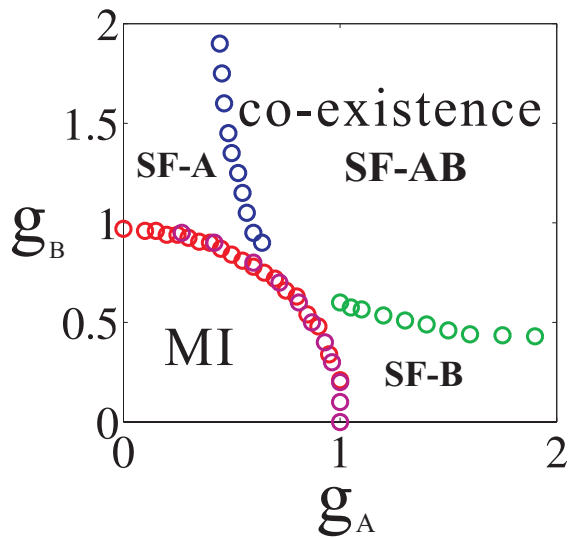}
\caption{Phase diagram on the parameter plane ($g_A$ and $g_B$) at zero temperature, i.e. for different atom-photon coupling constants (in the unite of frequency).
Four different phases are indicated in the plot. Other parameters used in the simulations are $\kappa=0.4$, $\kappa^{\prime}=0.2$, $\omega=2.7$, $\mu=0.2$, $\epsilon_A=2.7$, and $\epsilon_B=2.5$.}
\label{Fig:3}
\end{figure}

To verify above perturbative analyses, we solve the mean-field Hamiltonian in Eq. (\ref{mmf}) by using direct numerical simulations.
For the zero temperature, we solve the equations for different values of $g_A$ and $g_B$ by fixing the value of transition energy $\epsilon_{A(B)}$ and photon dispersion $\Omega_{+(-)}$.
Fig. \ref{Fig:3}  demonstrates the phase diagram of our system in the parameter plane $g_A$ and $g_B$ at zero temperature.
As conjectured by the perturbative analysis, we have the phase diagram for the QPT of light in our dual-species configuration, where the boundary for SF-A/B to SF-AB phases is turned into crossovers. Due to the Josephson-like coupling effect mentioned above. The co-existence SF-AB state in Fig. \ref{Fig:3} also occupies a much broaden area in the direct  numerical simulations.

\begin{figure}
\includegraphics[width=8.0cm]{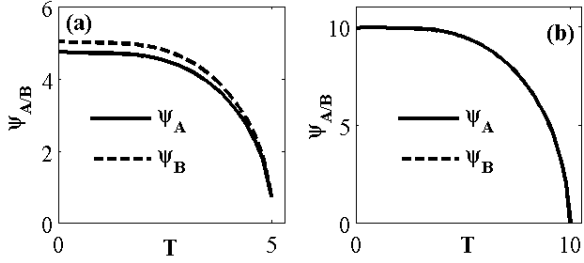}
\caption{Superfluid order parameters $\psi_A$ and $\psi_B$ versus
the temperature $T$ (in the unit of Kelvin) with different values of (a) $g_A=2.0$,
$g_B=0.1$; and (b) $g_A=2.0$, $g_B=2.0$. Other parameters used are
the same as those in Fig. \ref{Fig:3}. Here the hopping constants are $\kappa=0.4$ and $\kappa^{\prime}=0.2$.} \label{Fig:4}
\end{figure}
\begin{figure}
\includegraphics[width=8.0cm]{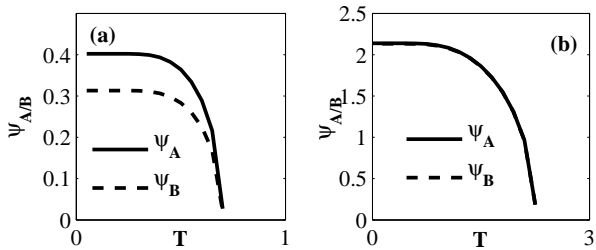}
\caption{Superfluid order parameters $\psi_A$ and $\psi_B$ versus the temperature $T$ (in the unit of Kelvin). All the parameters used are the same as those in Fig. \ref{Fig:4}, but with different hopping constants, $\kappa=0.35$ and $\kappa^\prime=0.175$.}
\label{Fig:5}
\end{figure}

For non-zero temperature, we plot in Fig. \ref{Fig:4}(a) and (b) the superfluid order parameters $\psi_A$ and  $\psi_B$ as a function of the temperature, $T$, with different values of $g_A$ and $g_B$, respectively. 
As shown in Fig. \ref{Fig:4}(a), for the parameters  $g_A=0.1$ and $g_B=2.0$, the system is in the SF-B phase at zero temperature.
A clear finite temperature insulator to superfluid transition is found at the critical temperature $T_c \approx 5$.
Notice that $\psi_B>\psi_A$ but the differences between $\psi_A$ and $\psi_B$ remains  small throughout the whole temperature range in the SF-B phase, indicating the importance of Josephson coupling.
For the co-existence SF-AB state at zero temperature, the two curves for $\psi_A$ and $\psi_B$ stay close  to each other for the whole temperature range.
The overall magnitudes of $\psi_A$ and $\psi_B$ are larger by a ratio of factor $2$ when compared with the case with $g_A=0.1$ and $g_B=2.0$. 

To examine more carefully the effect of Josephson-like coupling, we show again in Fig. \ref{Fig:5} the temperature dependence of $\psi_A$ and $\psi_B$ with all the same parameters except changing the hopping constants  to $\kappa^\prime=0.175$ and $\kappa=0.35$.
Comparing to the cases in Fig. \ref{Fig:4}, we find that the magnitudes of $\psi_A$ and $\psi_B$ become both smaller due to the reduced Josepshon coupling effect, and the difference between $\psi_A$ and $\psi_B$ becomes larger.
Notice that the values of $\Omega_{sym}$ and $\Omega_{ant}$ are increased by decreasing $\kappa^\prime$ and $\kappa$ (see equations (26), (27), (31), (32)) and the system is driven towards the MI phase.
In fact we find that $\psi_A=\psi_B=0$ and the system is already in the MI regime for $\kappa^\prime=0.2$ and $\kappa=0.3$ with our chosen set of parameters.

With the two-site Hamiltonian by including two different atomic species, we show the phase diagrams for the quantum phase transitions of light by applying the Bose-Hubbard Hamiltonian for photons with the Jaynes-Cummings photon-atom interaction. Using a self-consistent mean-field approximation, we analyze the equations for the superfluid order parameters analytically and numerically. Four different phases are found in our proposed system, including an isolated Mott-insulator phase, and three superfluid phases labelled as  SF-A, SF-B, and  co-existence SF-AB states. The transitions between the different superfluid phases are found to be smeared out by the Josephson-like coupling effect between different types of atoms. Our results demonstrate the  possibility to implement a photonic cavity system as a quantum simulator based on different atomic species. As the studies in condensed matter physics, more exotic phases and richer phase diagrams for the quantum phase transitions of light are expected with more complicated configurations and multiple atomic species in the proposed  model.

\acknowledgments
The authors are indebted to Andrew Greentree, Leon Soi, Chaohong Lee, and Sungkit Yip for helpful discussions. This research was supported by the National Science Council of Taiwan,  under the contract number of 98-2112-M-007-001-MY3.

\end{document}